\documentclass[twocolumn,showpacs,preprintnumbers,amsmath,amssymb]{revtex4}

\usepackage{graphicx}
\usepackage{dcolumn}
\usepackage{bm}

\def\bea{\begin{eqnarray}}
\def\eea{\end{eqnarray}}

\def\pp{\mbox{$p$-$p$} }
\def\auau{\mbox{Au-Au} }

\def\aa{\mbox{A-A} }
\def\nn{\mbox{N-N} }

\def\deta{$\eta_\Delta$ }

\begin{document} 

\preprint{Version 1.1}

\title{

Glasma flux tubes vs minimum-bias jets in 2D angular correlations on $\eta$ and $\phi$ 
}

\author{Thomas A. Trainor}
\address{CENPA 354290, University of Washington, Seattle, WA 98195}


\date{\today}

\begin{abstract}
Angular correlations measured in heavy ion collisions at the Relativistic Heavy Ion Collider (RHIC) include a same-side (SS) 2D peak which is strongly elongated on pseudorapidity $\eta$ in more-central \auau collisions. The elongated peak has been referred to as a ``soft ridge.'' The SS peak is consistent with expected jet correlations in peripheral \aa and p-p collisions. A saturation-scale argument has been proposed to explain the origin of the elongated SS peak  in terms of correlations from Glasma flux tubes interacting with radial flow. In this analysis we review the details of the proposed argument in comparison to perturbative QCD predictions of jet yields and correlations. We find that the proposal is inconsistent with several features of measured spectra and correlations.
\end{abstract}


\maketitle

 \section{Introduction}

Data from heavy ion collisions at the Relativistic Heavy Ion Collider (RHIC) have been interpreted within a hydro context to demonstrate formation of a thermalized, flowing partonic medium with small viscosity~\cite{qgp1,qgp2}. However, differential spectrum and correlation analysis reveals a jet-like contribution whose variation with \aa centrality and collision energy seems to contradict hydro expectations~\cite{axialci,hardspec,daugherity}. In particular, a large-amplitude 2D peak at the origin in angular correlations on ($\eta,\phi$) expected from jet formation persists even in central \auau collisions, albeit the peak is elongated on pseudorapidity $\eta$ relative to a nominally symmetric jet cone~\cite{axialci,daugherity}.

The elongated jet-like peak has recently been reinterpreted in terms of mechanisms other than parton scattering and fragmentation, including ``triangular flow'' resulting from initial-state transverse shape fluctuations (i.e.\ sextupole and other azimuth multipoles) which might in turn modulate radial expansion~\cite{gunther} and interaction of  ``Glasma flux tubes'' with radial flow~\cite{mg,gmm}. 

In this article we confront the latter strategy,  saturation-scale arguments based on Glasma flux tubes. We establish direct comparisons between saturation-scale predictions of particle production and measured yields. We examine a statistical argument for estimating the amplitude of the same-side peak in terms of Glasma flux tubes coupled with radial flow. And we compare the $\eta$ dependence of theory predictions with measured 2D angular correlations as a test of saturation-scale theory. We conclude that the Glasma flux-tube  description is falsified by spectrum and correlation data, whereas, perturbative QCD provides a quantitative description.

 \section{Analysis Method}

We relate saturation-scale arguments and associated correlation predictions directly to measured particle production and minimum-bias 2D angular correlations from \auau collisions at $\sqrt{s_{NN}} = 200$ GeV in the context of the Glauber model of \aa collision geometry.

\subsection{2D angular correlations}

Minimum-bias 2D angular correlations are constructed as autocorrelations on difference variables $\eta_\Delta = \eta_1 - \eta_2$ and $\phi_\Delta = \phi_1 - \phi_2$~\cite{inverse,ptscale,axialci,daugherity}. No ``trigger'' particle is used to define the event-wise angular origin. The only $p_t$ cut applied defines the lower limit of the momentum acceptance at $p_t = 0.15$ GeV/c. So-called forward-backward correlation measurements are 1D projections of 2D angular correlations onto $\eta_\Delta$ and so contain reduced information. Dihadron azimuth correlations are 1D projections onto $\phi_\Delta$, with similar information loss.

A per-particle correlation measure can be defined by 
\bea
\frac{\Delta \rho}{\sqrt{\rho_{ref}}} &=& \frac{\bar N}{\Delta \eta \Delta \phi} \, (r-1),
\eea
where $\bar N$ is the mean multiplicity in angular acceptance $(\Delta \eta,\Delta \phi)$ and histogram element $r_{ab}$ is the ratio of event-wise sibling pairs to mixed pairs in 2D bin $(a,b)$ within the pair angular acceptance. ${\bar N}(b)/{\Delta \eta \Delta \phi} = \rho_0(b)$ estimates the single-particle 2D angular density for impact parameter $b$ averaged over the acceptance. It is related to the mixed-pair reference by $\rho_0(b) \approx \sqrt{\rho_{ref}}$.

The observed correlation structure in 2D histograms consists of a few components modeled by simple functional forms. Within the nominal STAR TPC acceptance $(\Delta \eta,\Delta \phi) = (2,2\pi)$ the significant structure includes two $\eta_\Delta$-independent multipoles (dipole and quadrupole), a same-side (SS) 2D peak, and a 1D Gaussian on $\eta_\Delta$ having negligible amplitude in more-central \auau collisions. The SS 2D peak and away-side (AS) azimuth dipole have been interpreted in terms of minimum-bias jets (minijets) with typical parton energy $\approx 3$ GeV~\cite{fragevo,tom1,tom2}.

\subsection{Glauber model of \aa geometry}

The Glauber Monte Carlo model relates collision parameters $N_{bin}$ (\nn binary collisions) and $N_{part}$ (participant nucleons) to collision geometry measured by impact parameter $b$ and to a collision observable through cross section $\sigma(b)$ determined via the minimum-bias distribution on multiplicity $n_{ch}$ within some angular acceptance.

Centrality measure $\nu \equiv 2 N_{bin} / N_{part}$ is a convenient measure for studying correlations relative to a Glauber linear superposition (GLS) reference in which \aa collisions are assumed to be linear superpositions of \nn collisions. The two-component model of \aa collisions described in Ref.~\cite{kn}, when extrapolated from measured \pp collisions, is an example of a GLS reference.
As a context for saturation-scale arguments note that $R \sim N_{part}^{1/3} \sim \nu$, $\rho_{part} \sim N_{part} / R^2 \sim \nu$ and $N_{bin} \sim N_{part}^{4/3} \sim \nu N_{part}$.

 \section{Saturation-scale arguments}

Saturation-scale arguments applied to \aa collisions depend on longitudinal overlap of nucleon wave functions at small $x$~\cite{boris} and gluon transverse phase-space cell occupancy approaching unity. The result describes the distribution of gluons in a light-cone wave function produced by hard scattering off a large nucleus~\cite{mueller}.

\subsection{Saturation-scale gluon densities}

The problem addressed in Ref.~\cite{mueller} is early-time approach to thermalization, via multiple small-angle scattering in the transverse plane, of gluons released or freed from projectile nucleons near mid-rapidity. 
The saturation-scale treatment does not address large-angle (semihard) scattering of initial-state gluons. ``At early times there is a negligible transfer of longitudinal momenta into transverse momenta''~\cite{mueller}.
In that description gluons interact only with other gluons of the same rapidity during early-time  thermalization. 


In Ref.~\cite{mueller} $b$ represents a  transverse radius relative to the collision axis in central \aa collisions.
The quark-gluon color factor is $C_F = (N_c^2 - 1) / 2N_c = 4/3$. The strong-coupling constant is described by
${1}/{\alpha_s(Q^2)} \approx 0.7 \ln(Q^2 / \Lambda^2)$,
with  $\Lambda \sim 200$ MeV~\cite{eeprd}.
$\alpha_s C_F$ is the coupling strength of valence (projectile) quarks to soft gluons.
The projectile nucleon transverse density $\rho_N(b)$ integrates to A nucleons (for central collisions). The valence-quark density $3\rho_N(b)$ is the source for a field of soft gluons described in the low-density limit by
\bea \label{muel1}
\frac{d^2 N_g}{dy\, db^2} &=& \frac{dxG_A(x,Q^2)}{db^2} = 3\rho_N(b) \frac{\alpha_s C_F}{\pi} \ln(Q^2 / \mu^2) \\ \nonumber
&=& \rho_N(b) xG(x,Q^2).
\eea

At high gluon densities saturation should occur when $2C_F / \alpha_s$ gluons occupy each transverse phase-space cell. The saturation scale $Q_s^2$ is then defined by
\bea \label{muel2}
 \frac{dxG_A(x,Q_s^2)}{db^2} &=& c\,\frac{1}{\pi \alpha_s}\frac{C_F}{2\pi}  Q^2_s(x,b),
 \eea
where $c$ is the fraction of gluons released from projectile nucleons  in central \aa collisions. 
Whereas in the low-density limit $xG(x,Q^2)$ is slowly varying on $Q^2$, in saturation $xG(x,Q_s^2) \sim 1/\alpha_s(Q_s^2) \sim \ln(Q_s^2 / \Lambda^2)$. 


\subsection{Saturation-scale hadron production}

We now consider the \aa centrality dependence of saturation following Ref.~\cite{kn}. $b$ represents the \aa impact parameter, and $\rho_N(b) \rightarrow \rho_{part}(b)$ is the participant-nucleon transverse density. Combining Eqs.~(\ref{muel1}) and (\ref{muel2}) saturation sale $Q_s^2(x,b)$ is obtained by solving iteratively
\bea
 Q_s^2(x,b) &=& \frac{2\pi}{C_F}\pi\alpha_s(Q_s^2)\,  xG(x,Q_s^2)\, \rho_{part}(b).
\eea
Because  $\alpha_s(Q_s^2)\, xG(x,Q_s^2) \sim 1$ we expect $ Q_s^2(x,b) \sim \rho_{part}(b)$, and in Ref.~\cite{kn}  $Q^2_s \approx  2\rho_{part} / 3$ GeV$^2$. But $\rho_{part}(b) \approx \nu / 2$, where $\nu \equiv 2N_{bin} / N_{part}$~\cite{powerlaw}. So $Q^2_s \approx  \nu/ 3$ GeV$^2$, corresponding to the number of \nn binary collisions per participant pair.


Integrating Eq.~(\ref{muel1}) over $b^2$ (and $y \approx \eta$) we obtain
\bea
\frac{dN_g(x,Q_s^2)}{d\eta} = c\, xG(x,Q_s^2)\, N_{part}(b)
\eea
with $c \sim 1$. 
Assuming $dN_g/d\eta \approx (3/2)\, dn_{ch}/d\eta $ (LPHD)
\bea \label{kn5}
\frac{2}{N_{part}} \frac{dn_{ch}}{d\eta} &=& (4/3)\,  c\, xG(x,Q_s^2) \sim \frac{1}{\alpha_s(Q_s^2)}.
\eea
Given ${1}/{\alpha_s(Q_s^2)} \sim  \ln(Q_s^2 / \Lambda^2)$, $Q_s^2 \approx \nu / 3$ GeV$^2$ and $\Lambda \approx 0.2~\text{GeV}$ we obtain
\bea \label{kn6}
\frac{2}{N_{part}} \frac{dn_{ch}}{d\eta} &\approx& B \ln(8\, \nu)
\eea
where $B$ is an $O(1)$ constant determined from data~\cite{kn}. For comparison, the eikonal Glauber two-component form is (first line)
\bea
\frac{2}{N_{part}} \frac{dn_{ch}}{d\eta} &=& \rho_{pp}\left[ 1 + x(b)(\nu - 1)  \right] \\ \nonumber
&=& S_{NN} + \nu H_{AA}(b),
\eea
with $\rho_{pp} \approx S_{NN} + H_{NN}$, the peripheral limit of \aa collisions~\cite{kn,ppprd,hardspec}. Parameter $x(b) \approx H_{AA}(b) / (S_{NN} + H_{NN})$ 
(second line) actually has a strong centrality dependence, in contrast to the fixed $x$ value in Ref.~\cite{kn}. 

\subsection{Comparison with hadron production data}

\begin{figure}[h]
 \includegraphics[width=1.65in,height=1.6in]{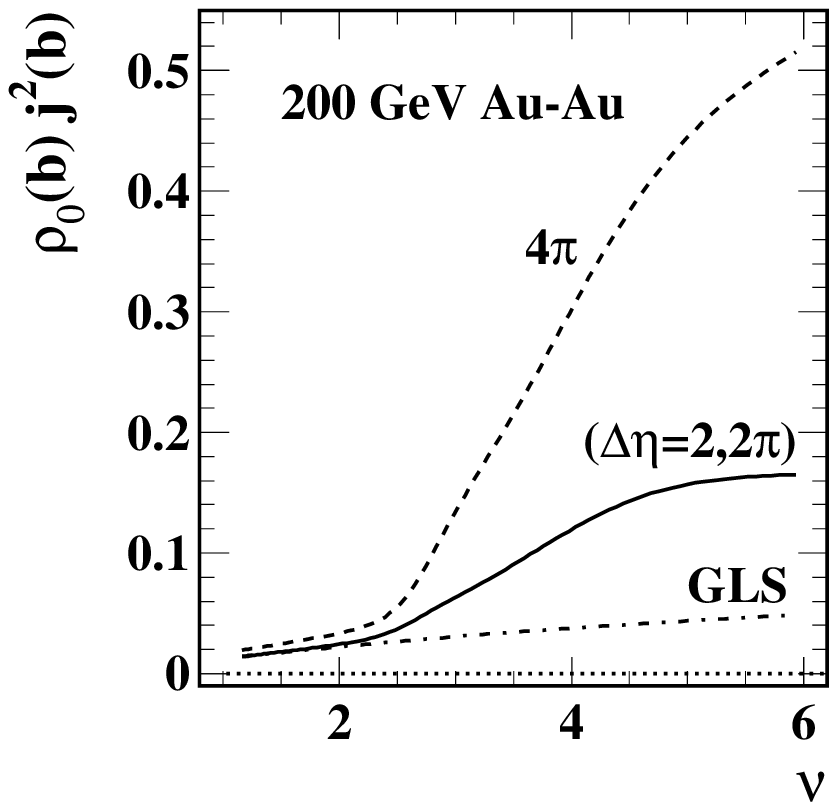}
   \includegraphics[width=1.65in,height=1.63in]{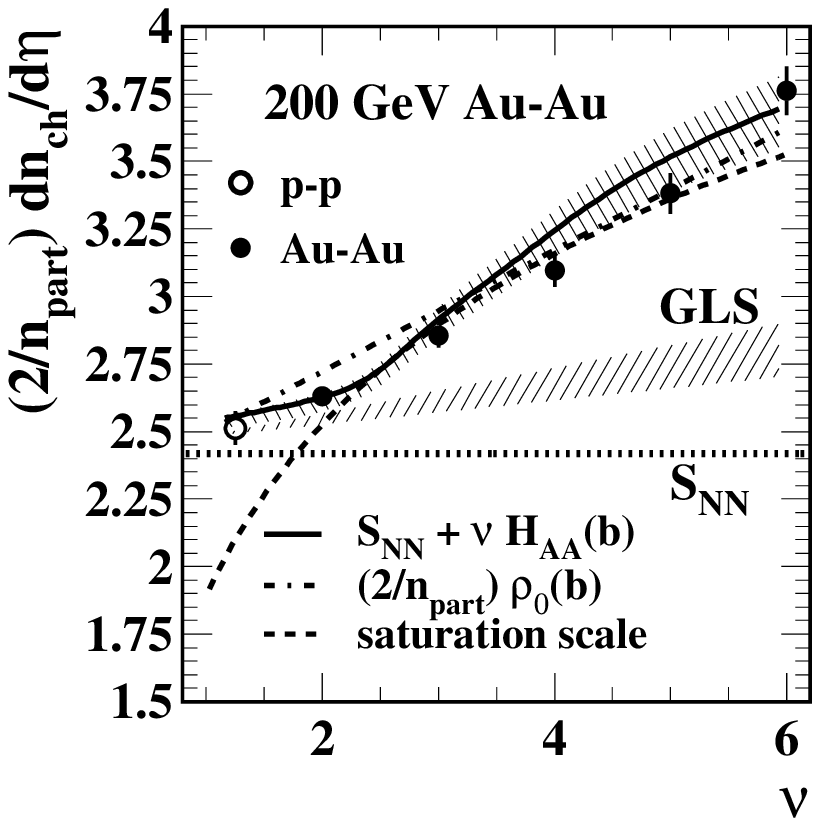}
\caption{\label{partcorr}
Left: Centrality dependence of the same-side 2D peak volume (solid curve) measured by pair ratio $j^2(b)$ within acceptance $\eta \in [-1,1]$ in combination with single-particle angular density $\rho_0(b)$. 
Right: Particle production trends vs centrality $\nu$ from spectrum data (points), from minijet angular correlations (bold solid curve), from the eikonal Glauber model of Ref.~\cite{kn} with $x = 0.09$ (dash-dotted line) and from a saturation-scale argument (bold dashed curve).
 } 
 \end{figure}

We can compare the saturation-scale prediction with measured particle production and angular correlation trends. In Fig.~\ref{partcorr} (left panel) the jet-correlated pair density (same-side 2D peak integral) is averaged over a limited angular acceptance (solid curve)~\cite{daugherity}. The dashed curve shows the result for $4\pi$ angular acceptance.
By invoking pQCD jet number $\bar N_j(b)$ per \aa collision appearing within the angular acceptance the corresponding jet fragment density can be inferred, represented by quantity $\nu H_{AA}(b)$~\cite{fragevo}.

Two-component particle yield model $(2/N_{part})\rho_0(b) = \rho_{pp}[1 + x(\nu - 1)]$, representing the eikonal model in Ref.~\cite{kn} with $x = 0.09$, is shown as the dash-dotted line in Fig.~\ref{partcorr} (right panel).  Solid curve $S_{NN} + \nu H_{AA}(b)$ includes $H_{AA}(b)$ inferred from two-particle jet correlations in Ref.~\cite{jetsyield} which increases by factor 4.5 from peripheral to central \auau collisions, factor 1.5 representing increase of the dijet cross section and factor 3 representing increase of the mean jet fragment multiplicity arising from fragmentation function modification~\cite{fragevo,jetsyield}.  The effective $x(b)$ value (slope on $\nu$) thus increases from $x \approx 0.02$ inferred directly from \pp spectra~\cite{ppprd} (trend labeled GLS) to $x \approx 0.1$ consistent with more-central \auau yields.

The bold dashed curve is saturation-scale trend $(2.5/2.25)\,0.82\ln(8\nu)$ from Eq.\ (19) of Ref.~\cite{kn} scaled from 130 to 200 GeV and using the equivalent of $Q_s^2 / \Lambda^2$ determined in the present analysis. Prefactor $B = 0.82$ established in Ref.~\cite{kn} matches 130 GeV  \auau data in more-central collisions. 
If the prediction is scaled to match more-central yield data (where saturation would be more likely) it fails dramatically for more-peripheral collisions. The prediction is concave downward, whereas the spectrum data (points~\cite{hardspec}) are significantly concave upward.

The two-component models in Fig.~\ref{partcorr} (right panel) assume soft component $S_{NN} \approx 2.4$ independent of centrality, any yield centrality dependence arising from parton scattering and fragmentation, whereas saturation-scale arguments predict a soft-component yield varying strongly with centrality as $\ln(8\nu)$ with no significant parton scattering contribution.  
The soft (non-jet) component can be divided conceptually into a non-saturation part dominating \pp particle production (conventionally described by string fragmentation and/or soft Pomeron exchange) and a conjectured saturation part depending strongly on centrality but with no prediction of absolute magnitude. 
Because addition of fixed soft component $S_{NN}$ to measured hard component $\nu H_{AA}(b)$ derived independently from jet correlation data matches the spectrum yields (points) we conclude that the flux-tube contribution is negligible even in central \auau collisions.



 \section{2D angular correlations}

Minimum-bias ($p_t$-integral) 2D angular correlations provide an essential reference system for any correlation analysis in which $p_t$ cuts  are imposed to study jet phenomenology (e.g. trigger-associated analysis) or azimuth quadrupole systematics. 2D correlation data are accurately described by a simple fit model including three principal model components: (a) a same-side 2D peak at the origin well approximated by a 2D Gaussian for all minimum-bias data, (b) an away-side ridge well approximated by AS azimuth dipole $[1 - \cos(\phi_\Delta)]/2$ for all minimum-bias data and uniform to a few percent on $\eta_\Delta$ (having negligible curvature), and (c) an azimuth quadrupole $\cos(2\phi_\Delta)$ also uniform on \deta to a few percent over the full angular acceptance. (a) and (b) together have been interpreted as minimum-bias jets or ``minijets''~\cite{fragevo}. (c) is conventionally interpreted to represent elliptic flow, a hydrodynamic phenomenon. The 2D data are described to a few percent of the major-feature amplitudes over the entire angular acceptance. 
Fit residuals typically contain no significant nonstatistical structure.


 \begin{figure}[h]
  \includegraphics[width=1.65in,height=1.5in]{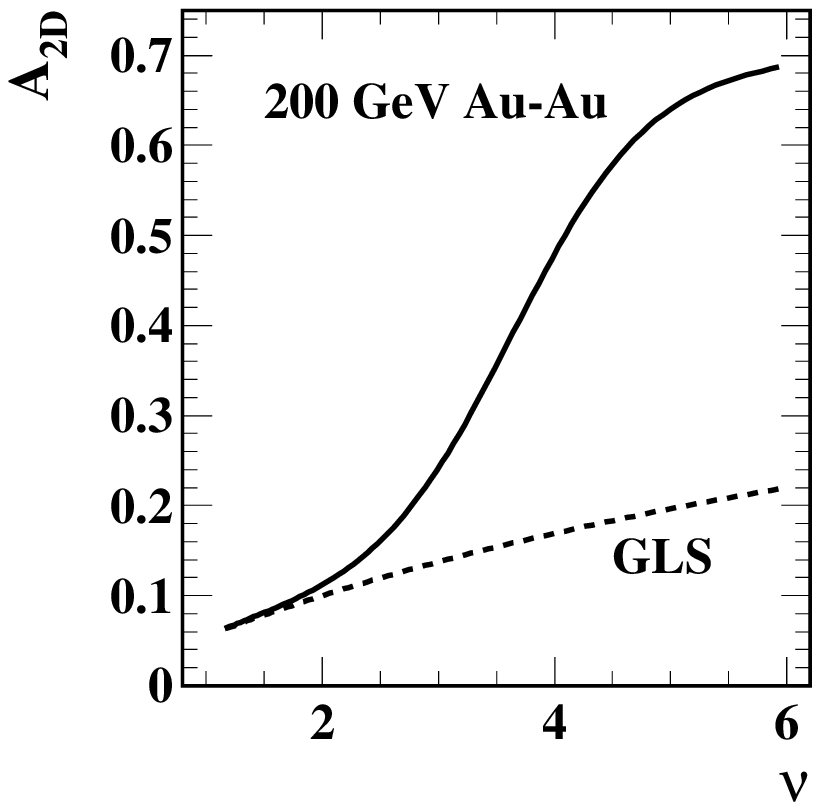}
  \includegraphics[width=1.65in,height=1.53in]{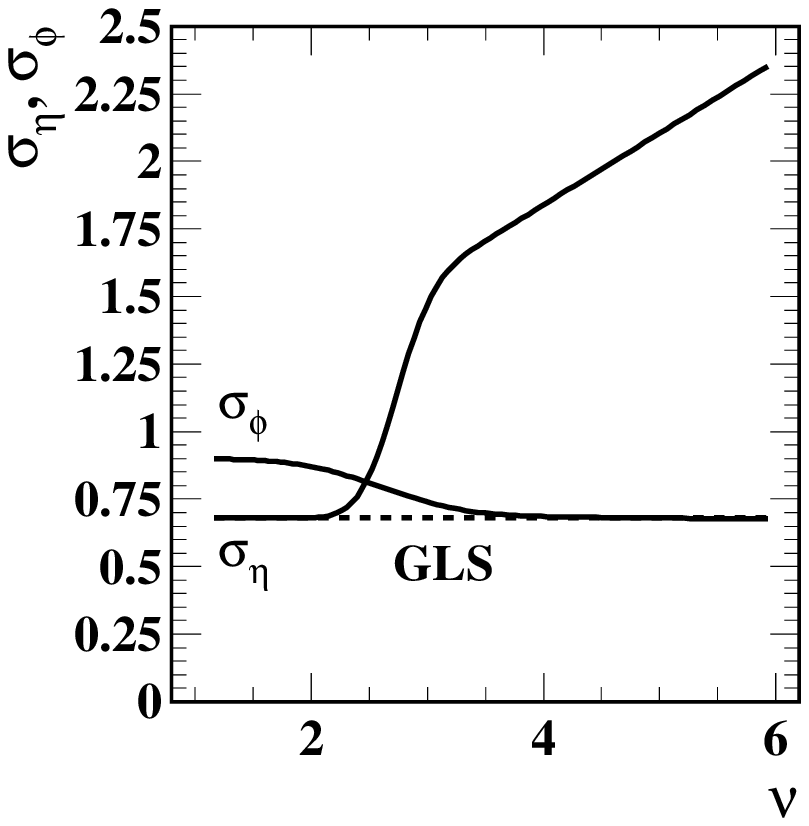}
\caption{\label{fitparams}
Left: Amplitude of a same-side 2D Gaussian fitted to minimum-bias 2D angular correlation data from 200 GeV \auau collisions~\cite{daugherity}.
Right: Fitted peak widths for the same-side 2D Gaussian. GLS indicates a Glauber linear superposition reference extrapolated from measured \pp collisions~\cite{ppprd}.
 } 
 \end{figure}

Figure \ref{fitparams} summarizes preliminary fitted SS 2D peak parameters vs centrality measure $\nu$ within nominal STAR TPC angular acceptance $(\Delta \eta,\Delta \phi) = (2,2\pi)$~\cite{daugherity}. $A_{2D}$ in the left panel is the fitted amplitude of the SS 2D Gaussian function. Its two r.m.s.\ peak widths are reported in the right panel. There is smooth variation with centrality, but a ``sharp transition'' in SS 2D peak properties occurs at a specific point on centrality---$\nu \sim 2.5$. The corresponding quadrupole data are reported in Ref.~\cite{davidhq}.

 \begin{figure}[h]
  \includegraphics[width=1.65in,height=1.5in]{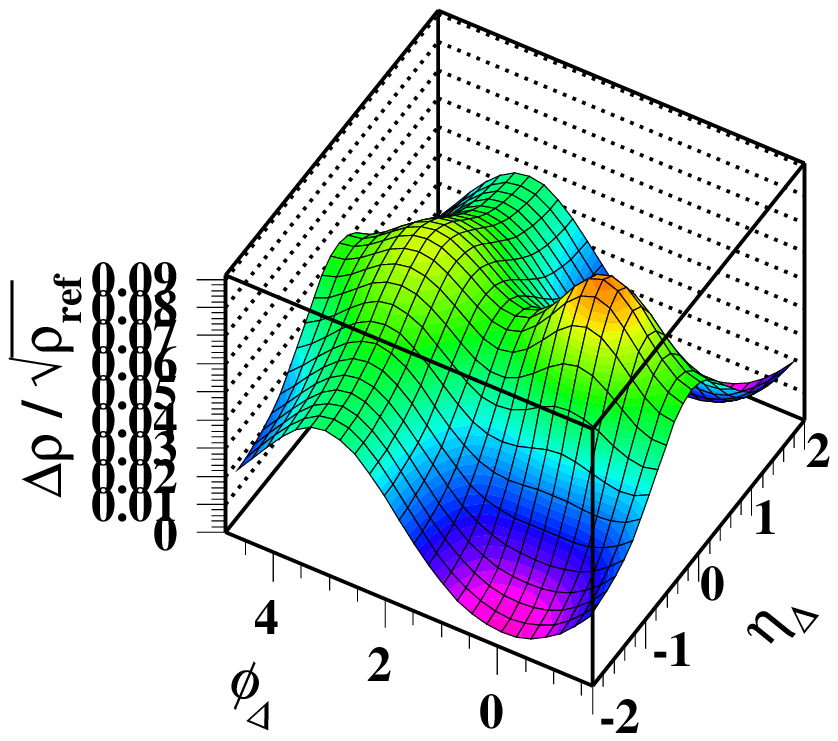}
  \includegraphics[width=1.65in,height=1.5in]{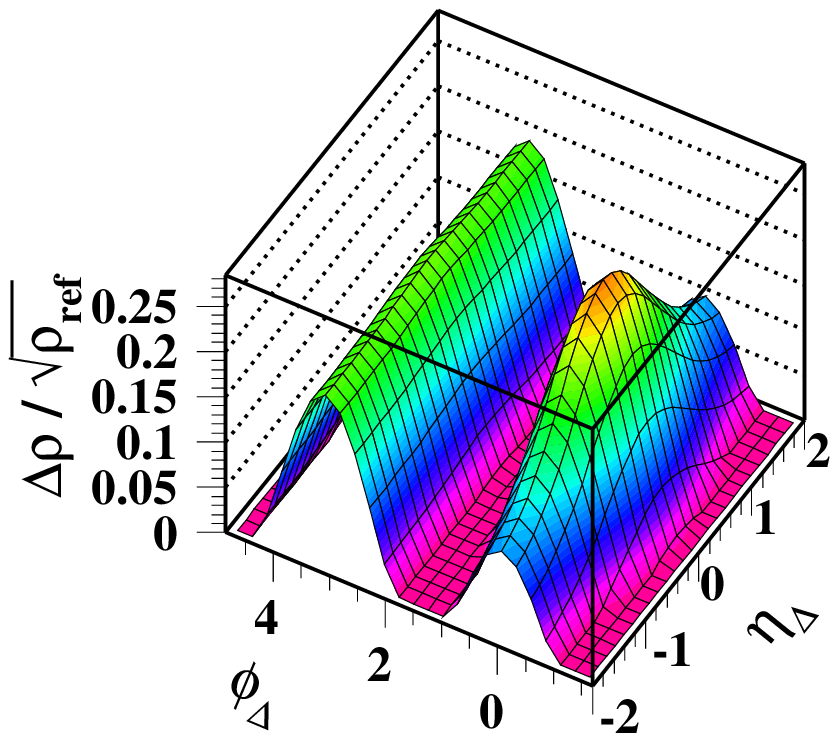}
 \includegraphics[width=1.65in,height=1.5in]{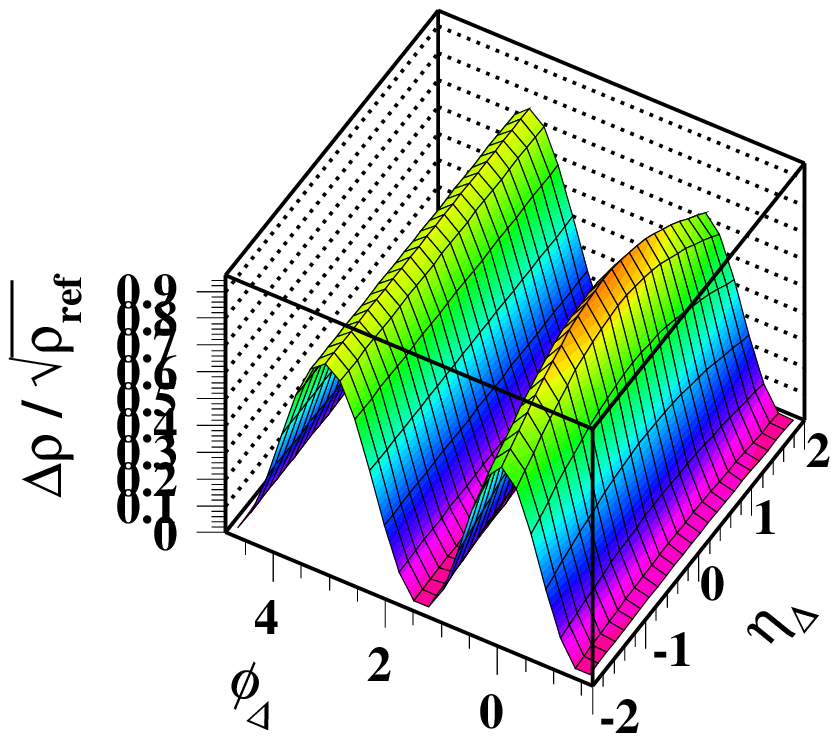}
  \includegraphics[width=1.65in,height=1.5in]{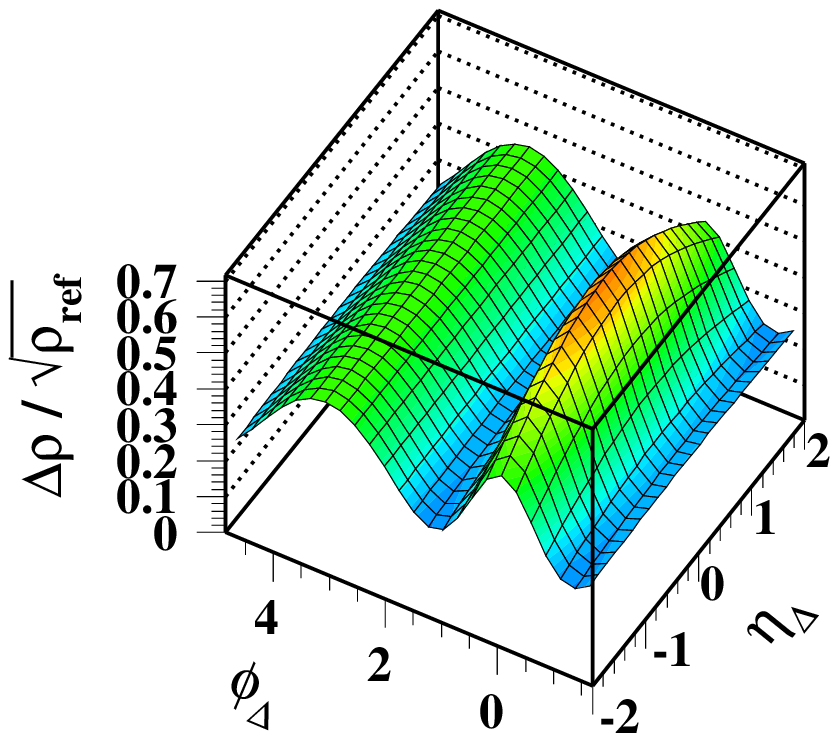}
\caption{\label{2dcorr} (Color online)
Angular correlation histograms for four centralities of 200 GeV \auau collisions based on fit parameters from Ref.~\cite{daugherity}. The centralities starting from upper left are given by $\nu = 1.25$ ($\sim$ \nn), 2.5, 4.5 and 6 ($b = 0$).
 } 
 \end{figure}

Figure \ref{2dcorr} shows examples of 2D angular correlations from four centralities of 200 GeV \auau collisions within acceptance $\eta \in [-1,1]$ based on the fit parameters of Ref.~\cite{daugherity}. The centralities correspond to $\nu = $ 1.25 ($\approx$ \nn collisions), 2.5, 4.5 and 6 ($b = 0$). The histograms are plotted within the STAR TPC angular acceptance usually adopted for 2D correlation analysis. 

Figure \ref{2dcorr} (upper right) with $\nu = 2.5$ (the transition point) includes a SS 2D peak with 1:1 aspect ratio (both peak widths $\approx 0.8$). Because the plotting format is $2\pi:4 \sim 3:2$ the symmetric peak seems to be elongated on $\eta$. Apparent peak elongation should be checked against the actual data (model fit) properties and plotting format. The SS peak seems to be superposed on a distinct ``ridge'' structure due to  apparent peak elongation and the underlying azimuth quadrupole contribution. Those data can be compared with Fig.\ 1 of Ref.~\cite{starridge}.

 \begin{figure}[h]
  \includegraphics[width=1.65in,height=1.6in]{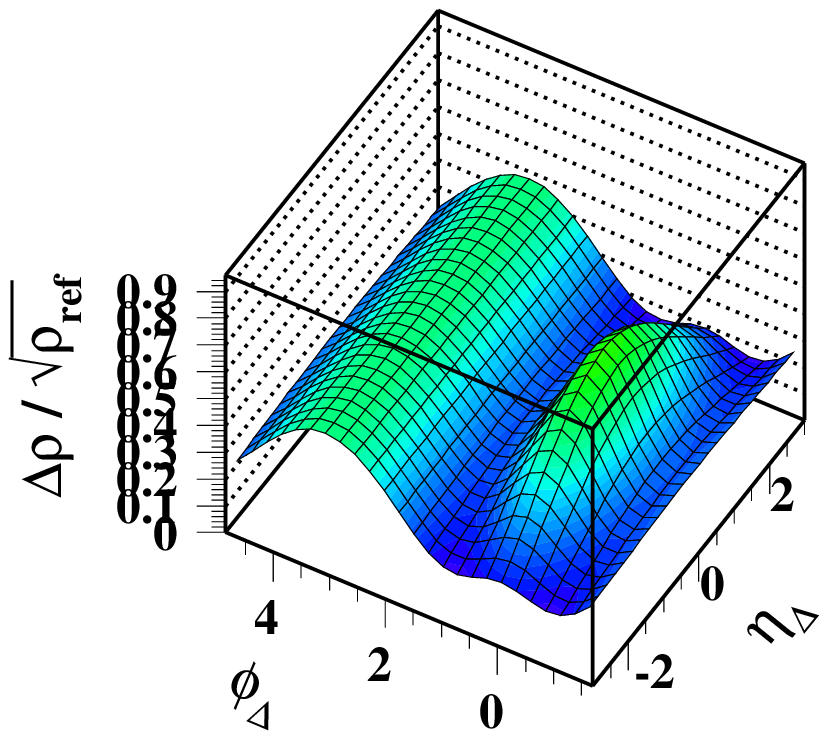}
  \includegraphics[width=1.65in,height=1.6in]{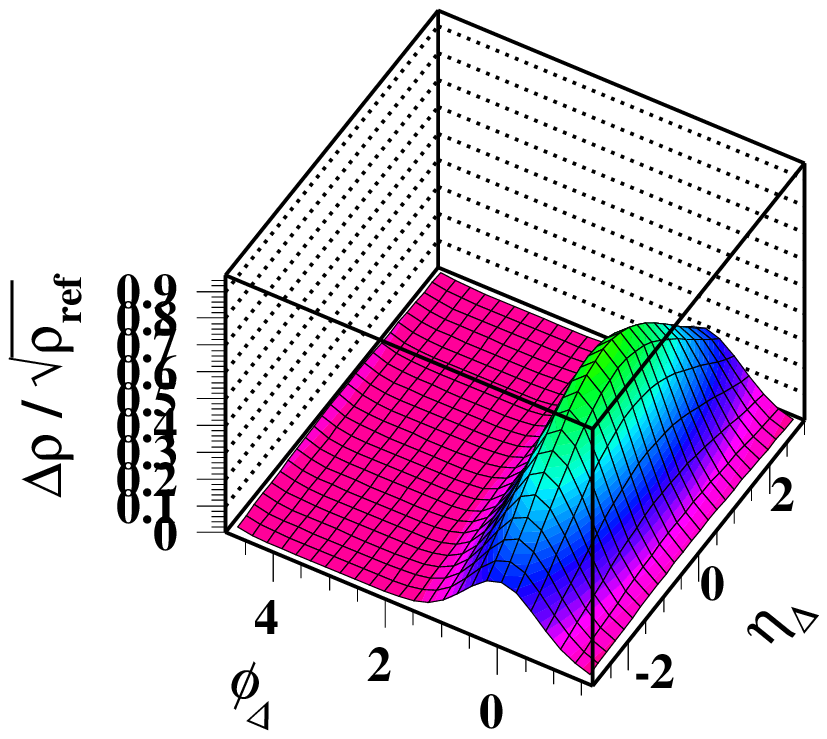}
\caption{\label{jetpeaks} (Color online)
Left: Angular correlations from 200 GeV \auau collisions with centrality $\nu = 4.5$ extrapolated to $\eta \in [-1.5,1.5]$ with nonjet fit components (quadrupole, 1D Gaussian on $\eta_\Delta$) subtracted to reveal nominal jet correlations.
Right: The previous histogram with the away-side dipole term subtracted to isolate the same-side 2D jet peak.
 } 
 \end{figure}

Arguments in Refs.~\cite{gmm,mg} pertain to the SS 2D peak at the angular origin. Figure \ref{jetpeaks} (left panel) illustrates nominal jet angular correlations in more-central \auau collisions ($\nu = 4.5$, lower-left panel in Fig.~\ref{2dcorr}) revealed when non-jet model components (mainly the azimuth quadrupole) are subtracted from 2D data histograms. The remaining elements are the SS 2D peak and the AS dipole. Figure \ref{jetpeaks} (right panel) shows the isolated SS 2D peak (dipole subtracted) extrapolated to larger $\eta$ acceptance  $\eta \in [-1.5,1.5]$, which is actually accessible within the STAR TPC albeit with increased systematic uncertainty. The SS peak for minimum-bias angular correlations is consistent with a single 2D Gaussian in more-central \auau collisions. SS peak structure retains a large curvature on $\eta_\Delta$ in all cases.

 \section{Glasma flux tubes and the SS peak}

Variation with \aa centrality of the amplitude and azimuth width of the SS 2D peak is described in terms of Glasma flux tubes, which are said to be ``pushed out'' by radial flow to form a SS 1D peak narrow on azimuth and uniform on $\eta_\Delta$. A statistical argument is introduced to express peak amplitude scaling in terms of  saturation scale $Q_s^2$.  A blast-wave model is invoked to determine the effects of radial flow on peak azimuth width and possibly amplitude variation.

\subsection{Statistics of a two-tiered hierarchy} \label{stats}

The treatment of correlations from Glasma flux tubes in Ref.~\cite{gmm} invokes a cluster or clan model~\cite{clan}. Such models apply as well to longitudinal string fragmentation~\cite{string} and transverse-scattered parton fragmentation~\cite{fragevo,jetsyield}.
Number fluctuations within angular acceptance ($\Delta \eta,\Delta \phi$) correspond to a running integral of angular correlations up to those limits~\cite{inverse}. ${\cal R}$ denotes a per-{\em pair} integral fluctuation measure comparable to correlation measure $r - 1$ defined in Ref.~\cite{axialci}. A corresponding per-{\em particle} integral measure is $\bar N {\cal R} \equiv \frac{\sigma^2_N - \bar N}{\bar N} = \Delta \sigma^2_N(\Delta \eta, \Delta \phi)$, where the last quantity is a {\em variance difference}---defined as a running integral (on angle bin size or scale) of angular correlations in the per-particle form $\Delta \rho(\eta_\Delta,\phi_\Delta) / \sqrt{\rho_{ref}}$~\cite{ptflucts,inverse}. The integral equation can be inverted to obtain differential number angular correlations within the angular acceptance from fluctuation scale dependence~\cite{inverse}. Inversion of mean-$p_t$ fluctuation scale dependence led to identification of minijet $p_t$ (as opposed to number) angular correlations as the source of those fluctuations~\cite{ptscale,ptedep}.

If gluon/hadron number $N$ results from event-wise production of $K$ sources, each emitting $n$ particles (both are random variables fluctuating about mean values) then
\bea \label{glasmeq}
\frac{\sigma^2_N - \bar N}{\bar N}  &=&  \frac{\sigma^2_n - \bar n }{\bar n} + \lambda \frac{\sigma^2_{n_1 n_2}}{\bar n} + \bar n\, \left[\frac{\sigma^2_K - \bar K}{\bar K}\right] + \bar n,
\eea
illustrating the additivity of per-particle measures. That expression could describe fragmentation of flux tubes, prehadrons from a Lund string or large-angle scattered gluons. 
The first term is a variance difference representing the integral of intra-source correlations---the SS 2D peak. The second term is a covariance representing the integral of source-source correlations---the AS ridge. $\lambda$ measures the fraction of sources (e.g.\ jets) with a single partner in the $\eta$ acceptance, anticipating a back-to-back jet interpretation. The third term represents event-wise source-number fluctuations due to \aa centrality fluctuations and centrality fluctuations of \nn collisions within \aa collisions. The fourth term represents a contribution due to the two-tiered hierarchy mechanism alone,  present even when sources and fragments are Poisson distributed. The third and fourth terms represent correlation contributions uniform on angle, not resolved peak structure.

In Ref.~\cite{gmm} the first and third terms of Eq.~(\ref{glasmeq}) are assumed to be negligible (Poisson processes), and the covariance term is not acknowledged. The fourth term $\bar n$ is attributed to the SS 2D peak amplitude. $\bar N {\cal R} \rightarrow \bar n$ or ${\cal R} \sim 1/\bar K$, with $\bar K = Q_s^2R^2 \sim N_{part}$: The mean number of flux tubes (sources) is proportional to the number of participant nucleons. Since $\bar N \sim [1/\alpha_s(Q_s^2)]\, Q_s^2R^2$, $(d\bar N/d\eta) {\cal R} \sim \bar n \sim 1/\alpha_s(Q_s^2) \sim \ln(8\nu)$ is taken as a measure of the mean per-particle 2D peak amplitude.
That is also the trend claimed for the per-participant-pair particle yield $(2/N_{part})\, dn_{ch}/d\eta$ in Ref.~\cite{kn}, as in Eq.~(\ref{kn5}). Because the fourth term of  Eq.~(\ref{glasmeq}) describes a uniform background contribution to angular correlations it is not relevant to the SS 2D peak.

In a jet interpretation of Eq.~(\ref{glasmeq}) the {first} term on the RHS measures intra-jet correlations---the SS 2D peak integral (not the amplitude), which is the actual subject of Ref.~\cite{gmm}. The total number of correlated pairs is $\bar K(\sigma^2_n - \bar n) \rightarrow  \bar N_j(b)\, n_{ch,j}^2(b)$, a product of the mean event-wise jet number and the number of fragment pairs per jet. A pQCD calculation of $\bar N_j(b)$ combined with measured angular correlations from Ref.~\cite{daugherity} quantitatively describes single-particle spectrum hard-component yields attributed to jets, as in Fig.~\ref{partcorr} (right panel)~\cite{jetsyield}.
Figure~\ref{jetpeaks} (left panel) shows two of the four elements contributing to $\bar N {\cal R} $ in Eq.~(\ref{glasmeq}). The integral of the SS 2D peak is represented by the first term in the RHS. The AS dipole peak integral is represented by the second term.



\subsection{Glasma flux tube centrality trend}

In Ref.~\cite{gmm}
the SS 2D peak amplitude for \aa collisions is defined by
\bea \label{eqmg}
\frac{\Delta \rho}{\sqrt{\rho_{ref}}} &=& \kappa\, \alpha^{-1}_s[Q_s(\sqrt{s_{NN}},b)]\, F(\phi_\Delta;\sqrt{s_{NN}},b),
\eea
where $F(\phi_\Delta)$ is a unit-normal peaked function centered at zero azimuth difference, its width $\sim 0.65$ varying only slightly with energy and centrality. 
The amplitude of $F(\phi) \propto 1/\sqrt{\sigma_{\phi_\Delta}}$ can then vary by at most 20\% given the slowly-varying trend of $\sigma_{\phi_\Delta}$ in Fig.\ 2 of Ref.~\cite{gmm}. However, $F(\phi_\Delta)$ is attributed to the dashed curve in Fig.\ 1 of Ref.~\cite{gmm} which varies by more than a factor 3.

In Ref.~\cite{eeprd} the strong-coupling constant is given by
\bea
1/\alpha_s(Q_s) &\approx& 0.7 \ln(Q_s^2 / 0.04~\text{GeV}^2),
\eea
 with $Q_s^2 \approx \nu / 3$ GeV$^2$ from Ref.~\cite{kn}. Thus, the SS 2D peak amplitude is predicted to vary as $\ln(8 \nu)$, just as for saturation-scale particle production in Eq.~(\ref{kn6}). The peak amplitude prediction for 200 GeV \auau in Fig.\ 1 (upper panel, solid curve) of Ref.~\cite{gmm} varies between 0.05 and 0.75, increasing by a factor 15 over $1< \nu < 6$. But $\ln(8\times 6) / \ln(8) = 1.9$. The solid curve from Ref.~\cite{gmm} then seems to be incompatible with Eq.~(\ref{eqmg}). 

The $\alpha_s^{-1}$ factor in Eq.~(\ref{eqmg}) is derived from an argument based on fluctuation measurements which would relate to the volume of a bounded 2D correlation peak, not the amplitude of a 1D peak on azimuth unbounded on $\eta$. 



\subsection{Glasma flux tube correlation description} \label{glascorr}

\begin{figure}[h]
   \includegraphics[width=1.65in,height=1.6in]{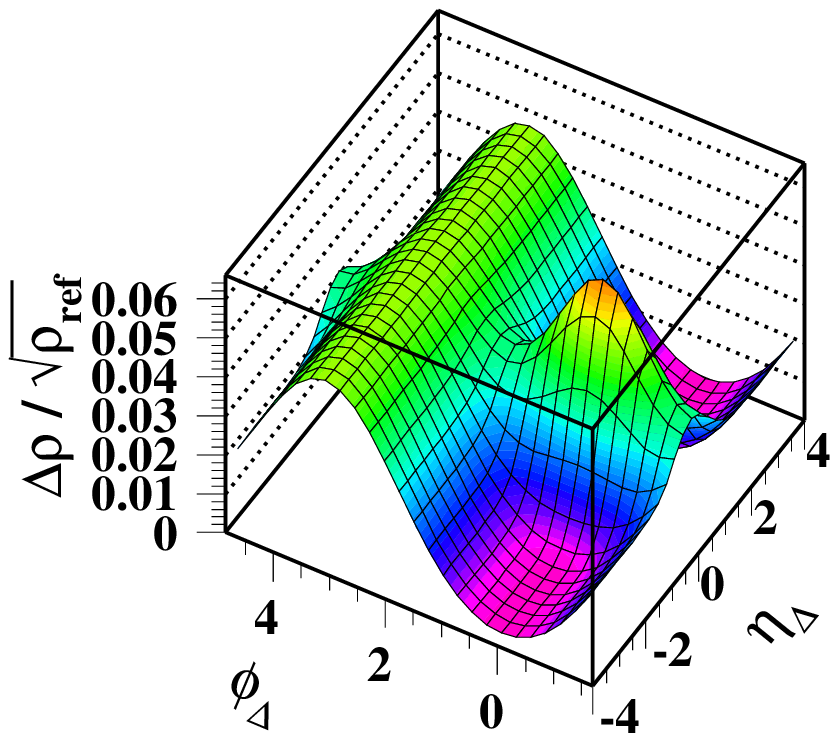}
 \includegraphics[width=1.65in,height=1.6in]{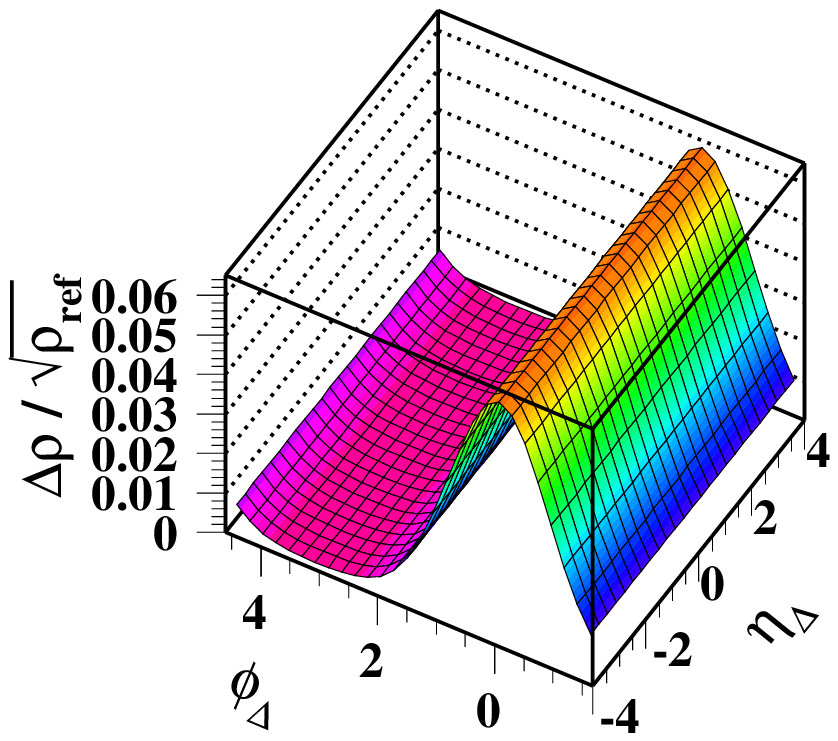}
 \includegraphics[width=1.65in,height=1.6in]{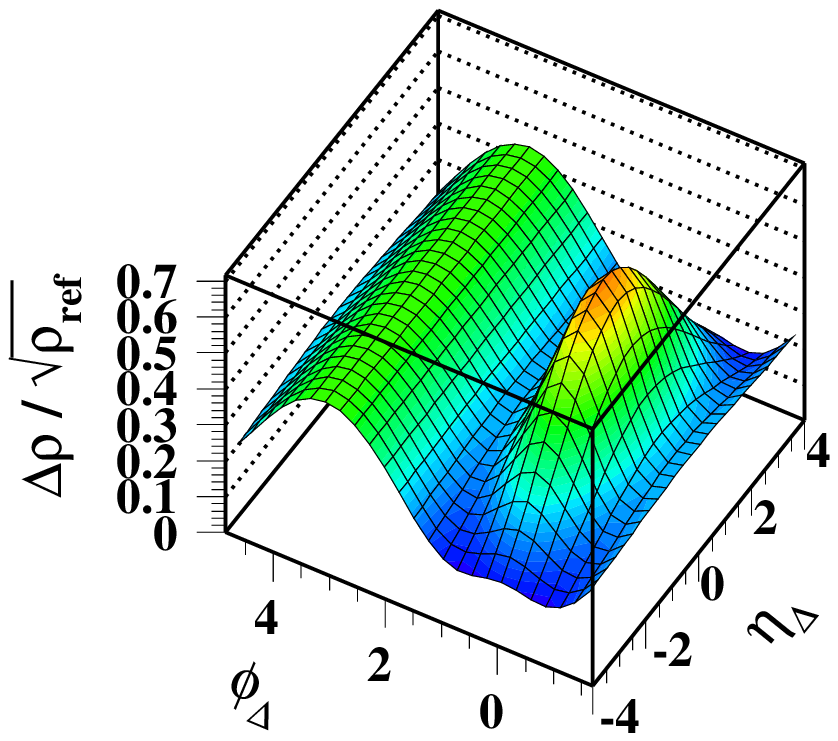}
  \includegraphics[width=1.65in,height=1.6in]{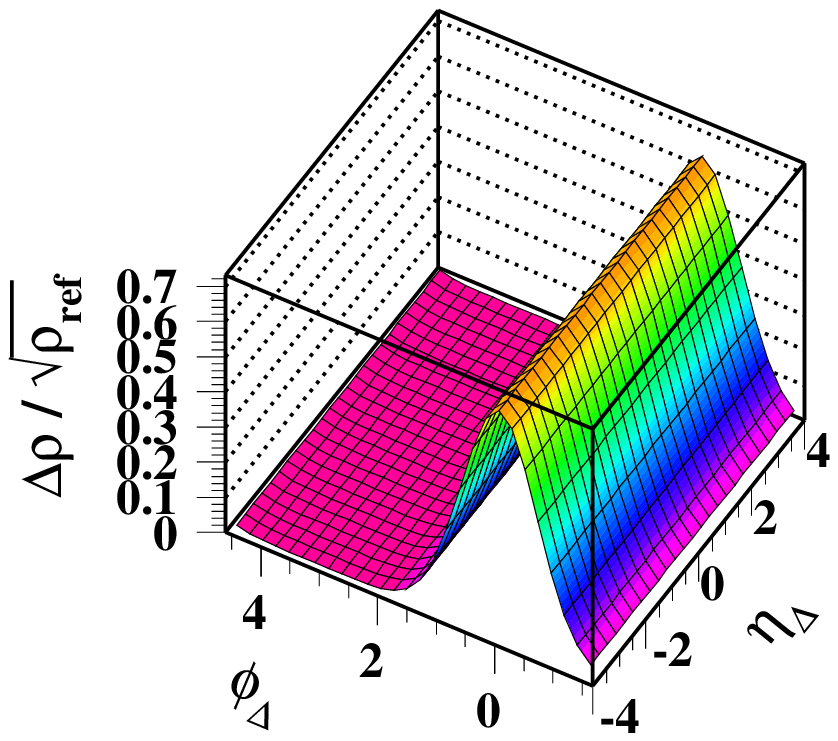}
\caption{\label{glasma} (Color online)
Comparison of fits to 2D data histograms (left panels) with Glasma flux tube predictions (right panels) for \auau collisions with $\nu = 1.25$ ($\approx$ \nn collisions, upper panels) and $\nu = 6$ ($b = 0$, lower panels).
} 
 \end{figure}

Figure \ref{glasma} provides a direct comparison between the full flux-tube correlation prediction and measured data. In Figure \ref{glasma} (upper panels) the flux-tube prediction is compared to peripheral \auau $\approx$ \pp angular correlations ($\nu = 1.25$ in Fig.~\ref{2dcorr}) extrapolated to $\eta$ acceptance $\eta \in [-2,2]$. The 1D Gaussian component on $\eta_\Delta$ has been removed, leaving SS and AS jet contributions. Evaluating the flux-tube ridge structure as a possible component of \pp correlations we conclude that the 2D data impose a small upper limit on ridge structure consistent with zero.

 In Figure \ref{glasma} (lower panels) the SS 2D peak in data from central \auau collisions ($\nu = 6$, the quadrupole is negligible) is substantially elongated on $\eta_\Delta$. However, the 2D data (left panel) do not require a significant flux-tube ridge structure (right panel). Although some triggered 2D angular correlations seem to suggest a separate ``ridge'' extending beyond a localized and symmetric 2D ``jet'' peak~\cite{starridge} the selected particle pairs are a subset of minimum-bias data which do not so indicate. If $p_t$ cuts are imposed significant deviations from an ideal Gaussian shape should be expected due to local charge, flavor, baryon-number and momentum conservation during parton fragmentation. For instance, SS peak shapes for like-sign and unlike-sign pairs may be quite different~\cite{axialcd}.

 \section{Discussion}

The principal goal of Refs.~\cite{gmm,mg} is reinterpretation of minijet phenomenology (e.g.\ systematics of SS 2D peak in angular correlations) in terms of a non-jet mechanism based on Glasma flux tubes. We review major elements of the Glasma flux-tube argument in comparison with a pQCD description of minijet manifestations in spectra and correlations.

\subsection{Ridge terminology and $\eta$ dependence}


Imposition of the term ``ridge'' on SS 2D correlation structure has caused significant confusion. The SS correlation structure is a single 2D peak under almost all conditions, although some $p_t$ cut conditions produce significant deviations from an ideal Gaussian shape. The combined terms ``soft'' and  ``ridge'' seem to distance SS correlation structure from a nominally-ideal jet phenomenon manifested by high-$p_t$ (hard) particles falling within a symmetric jet cone. 

The implied dichotomy of a hard (jet) peak and soft (non-jet) ridge is questionable. Most parton fragments in pQCD jets appear below 2 GeV/c~\cite{eeprd,fragevo}, just where thermalization and hydro flows are commonly believed to dominate RHIC collisions~\cite{nohydro}. Such ``soft'' fragmentation is susceptible to strong angular deformation depending on the collision system. The SS jet peak in \pp collisions is strongly elongated (2:1) in the azimuth direction~\cite{porter2} in contrast to comparable elongation (3:1) in the $\eta$ direction in central \auau collisions~\cite{porter2,axialci,daugherity}. 

As noted in Sec.~\ref{glascorr}, accurate measurement {and description} of correlation $\eta$ dependence is essential for correct physical interpretation. $\eta$ dependence permits separation of 2D correlations into $\eta$-independent (negligible curvature on $\eta_\Delta$) and $\eta$-dependent (strong curvature on $\eta_\Delta$) structure within the STAR TPC acceptance. 

\subsection{Glasma flux tube correlation mechanism}

According to Refs.~\cite{gmm,mg} early-stage rapidity correlations combined with late-stage radial flow produce a SS correlation structure elongated on $\eta_\Delta$. Early-stage correlations are attributed to fragmentation of Glasma flux tubes to separate gluons. Late-stage radial flow results from thermalization of the same flux-tube fragments. Both the proposed correlation mechanism and persistence of any such correlations in the measured final state are questionable for four reasons:

\paragraph{\bf Absence of conjectured initial correlations}

The concept of (color) correlation length seems to be confused with isolated flux tubes acting as distinct particle sources. In the case of isolated ``string'' fragmentation in \pp collisions (flux tube diameter $\sim$ hadron size) described by the Lund model we observe angular correlation structure {\em narrow} on $\eta_\Delta$ and nearly uniform on $\phi_\Delta$~\cite{porter1,porter2}. That (soft component) correlation structure falls to zero amplitude by mid-centrality in \auau collisions~\cite{daugherity}. The concept of correlated gluon emission from individual Glasma flux tubes is a theoretical conjecture unconfirmed by experimental evidence.

\paragraph{\bf Time sequence: boost vs emission}

According to Ref.~\cite{gmm} particles (gluons) resulting from flux tube fragmentation thermalize and produce radial pressure gradients which drive radial flow. The radial flow then boosts the correlated particles to produce narrow structures on azimuth. We might expect a large correlation effect when particles are emitted from a common source already boosted, in which case isotropic emission in the boost frame should appear as a directed jet of particles in the lab frame. A notable example is QCD jets.

If particles are emitted isotropically from a source stationary in the lab and then boosted radially, some pair opening angles will increase while others decrease. We should expect a minimal correlation effect. Thus, the time sequence is critical. For the sequence described in Ref.~\cite{gmm}, even if flux-tube fragments were initially strongly correlated in configuration space the thermalization process (which would also tend to reduce initial-state correlations) followed by late radial boost should result in negligible momentum angular correlations.

\paragraph{\bf Absence of radial flow} Radial flow is inferred from $p_t$ spectra by blast-wave fits within limited $p_t$ intervals below 2 GeV/c, based on the assumption that the entire spectrum in that $p_t$ interval corresponds to emission from a thermalized bulk medium. However, differential analysis of  200 GeV \auau $p_t$ spectra for  two hadron species reveals that spectrum evolution is dominated by the spectrum hard component, with most-probable $p_t \approx 1$ GeV/c~\cite{hardspec}.  Blast-wave parameters inferred from spectrum fits correspond closely to fragment distributions predicted quantitatively via pQCD~\cite{fragevo,nohydro}. 

\paragraph{\bf Central-limit attenuation}

According to the argument in Ref.~\cite{gmm} almost all final-state hadrons come from flux tubes. The number of flux tubes in central \auau collisions is $Q_s^2 R^2\approx N_{part} \approx 200$. Flux tubes at different radii should suffer different radial boosts and thus correspond to different peak widths. All such contributions would then be superposed event-wise and randomly distributed on the $2\pi$ azimuth acceptance. Superposition should lead to what can be termed central-limit attenuation: the many contributions sum to a near continuum on $\phi$ with small-amplitude fluctuations.

\subsection{Glasma flux tubes compared to QCD jets}

According to Ref.~\cite{gmm} Glasma flux tubes effectively replace minimum-bias jets as the fundamental QCD process in nuclear collisions. The statement ``Flux tubes arise naturally in...high energy collisions'' does not differentiate between Lund strings well-studied in elementary hadronic collisions and conjectured Glasma flux tubes. 

Flux tubes and large-angle scattered partons do share some formal similarities. 
Both objects may fragment to daughter gluons which then convert to hadrons according to LPHD. Both might result in SS angular correlations among detected hadrons. Both are nominally described by QCD, one in the perturbative limit, the other in the continuum limit. Both might involve transport from longitudinal to transverse momentum space. But  the two possess very different systematics which disfavor the former in comparisons with data. Minimum-bias jets conform to all aspects of pQCD in spectrum and correlation measurements except $\eta$ broadening, whereas Glasma flux tubes conform to almost no measurement features except the $\eta$ broadening which motivated their introduction.

\setcounter{paragraph}{0}

\paragraph{\bf Glasma flux tubes} 

In this one-component model the (soft) component arises from longitudinal fragmentation of projectile nucleons (via flux tubes). Large-angle parton scattering and fragmentation do not play a significant role. Flux tubes are transverse phase-space cells in an $\eta$-independent theory. The single parameter of the theory is the saturation scale $Q^2_s \propto \nu$. $1/\alpha_s(Q^2_s)$ is said to account for all $\sqrt{s}$ and centrality dependence.
  
The multiplicity of flux tubes is $\bar K \sim Q_s^2 R^2 \propto N_{part}$. The mean fragment multiplicity per flux tube is $\bar n \sim 1/\alpha_s(Q_s^2) \sim \ln(Q_s^2 / \Lambda^2) \sim \ln(8\nu)$. The total gluon/hadron multiplicity from all flux tubes in an \aa collision is $\bar K \bar n \sim Q_s^2 R_A^2 \ln(Q_s^2 / \Lambda^2) \sim N_{part} \ln(8\nu)$. The per-particle SS correlation amplitude is predicted by $(dn_{ch}/d\eta) {\cal R}  = \bar n \sim \ln(8\nu)$. There is no AS peak on azimuth in the theory---no tube-tube correlations, no momentum transfer between flux tubes. 

The flux-tube system relies on conjectured radial flow to  translate longitudinal momentum to transverse phase space (via thermalization and radial pressure gradients). Inference of radial flow from spectra relies on interpreting parton fragmentation as a hydro manifestation~\cite{hardspec,nohydro}.

\paragraph{\bf Minimum-bias jets}

In the two-component model~\cite{kn,ppprd} participant nucleons fragment longitudinally to fixed hadron multiplicity (Lund model) independent of \aa centrality (soft component). The contribution from minimum-bias jets (minijets) increases at least as fast as $N_{bin}$, rising to 1/3 of the total $n_{ch}$ in central \auau collisions at 200 GeV~\cite{jetsyield} (hard component).

The multiplicity of minimum-bias ($\approx 3$ GeV) jets in acceptance $\Delta \eta$ is $\bar N_j = N_{bin} \Delta \eta f(b)$, where \nn jet frequency $f(b)$ is slowly varying with centrality~\cite{eeprd,jetsyield}. The fragment multiplicity per jet in \pp collisions is $n_{ch,j} \sim y_{max}^2 \sim [\log(Q/m_\pi)]^2$~\cite{eeprd}, where $Q \approx 6$ GeV is the {dijet} energy, and increases to three times that number in central \auau collisions, consistent with strong modification of fragmentation functions without significant {\em net} energy loss from jets~\cite{fragevo}. 
Total multiplicity $n_{ch}$ in \aa collisions is the sum of soft and hard components. 

For angular correlations the per-particle SS 2D peak {\em volume} is measured by $\bar N {\cal R} = \bar N_j\, n_{ch,j}^2 /  n_{ch}$~\cite{jetsyield}. Angular correlations also include a prominent AS 1D peak on azimuth (AS ridge) whose centrality variation ($\propto N_{bin}$) closely matches the SS 2D peak and is very different from that expected from global momentum conservation ($\propto N_{part}$). AS ridge characteristics require large-angle scattering of parton pairs, with each parton localized on $x$ and with a broad distribution on difference $x_1 - x_2$, not longitudinally co-moving as assumed in Ref.~\cite{mueller}.

A specific energy dependence is expected for jet phenomena $\propto \log\{\sqrt{s_{NN}} / 13.5~\text{GeV}\}$  (inferred from $v_2\{2D\}$ data obtained from fits to 2D angular correlations~\cite{davidhq}). According to that trend the 62 GeV SS 2D peak amplitude should be a fraction 0.57 of the 200 GeV peak amplitude for \auau collisions, and that is observed~\cite{daugherity}.

\paragraph{\bf Comparison summary}

The minijet description has good predictive power for hadron spectrum and correlation structure. The spectrum hard component is predicted quantitatively in terms of minimum-bias fragmentation by folding a pQCD dijet spectrum with measured fragmentation functions (modified in more-central \aa collisions).~\cite{hardspec,nohydro}. The spectrum soft component may be a universal feature of all collision systems~\cite{nohydro}.

Event-wise reconstruction of jets in \pp collisions has been achieved down to 3-4 GeV by UA1~\cite{ua1} as well as STAR~\cite{starpp}. Minijet systematics in \auau collisions inferred  from particle production, spectra and minimum-bias correlations agree quantitatively with results from jet reconstruction in terms of (a) no jets lost to thermalization, (b) particle-number suppression at larger $p_t$ counterbalanced by larger enhancement at smaller $p_t$ corresponding to (c) nearly conserved total jet energy. Energy is rearranged within jets and angular correlations are modified by SS peak elongation on $\eta$, in contrast to elongation on $\phi$ for \pp collisions. 

The description based on Glasma flux tubes fails to provide a comparably accurate and comprehensive description. 
Theories which ignore 2D correlation structure on $\eta$ (e.g. curvatures), which deal only with projections onto azimuth, and which do not account for differential $p_t$ spectrum structure and its evolution with centrality and collision energy are not adequate to describe the ensemble of experimental results now available from RHIC.

 \subsection{Flux tubes, causality and cosmology}

It is argued in Ref.~\cite{gmm} that ``ridge'' correlations extending over large $\eta$ intervals require a source established at early times.  ``Correlations over several rapidity units can only originate at the earliest stages  of an ion collision when the first partons are produced. ...causality limits such effects [hydro modifications] to a horizon of roughly from one to two rapidity units.'' 
%
The argument can be summarized as (a) ``Flux tubes arise naturally in...high energy collisions.'' (b) Glasma flux tubes form very early in the collision ($< 1$ fm/c) (c) long-range correlations require an early-time source (causality argument), therefore (d) long-range correlations must be produced by Glasma flux tubes. 

Regarding (a,b) -- Two types of flux tubes can be distinguished. The flux tubes described in [13] of Ref.~\cite{gmm} (e.g.\ Lund model) are indeed expected in QCD but have a nonperturbative transverse size measured by $1/\Lambda_{QCD} \approx 1$ fm, whereas conjectured Glasma flux tubes are said to have a perturbative transverse size measured by $ 1/Q_s \approx 0.1-0.2$ fm. The latter are not an inevitable consequence of QCD as implied by Ref.~\cite{gmm}.

Regarding (c,d)  -- $\eta$ is a measure of polar angle.  Fragmentation extending over large time intervals can produce significant yields at polar angles far from $\pi / 2$ and hence at large $\eta$, especially with the large increase of small-$p_t$ jet fragments in more-central \auau collisions. 

An alternative broadening mechanism based on conventional pQCD is color connection of the scattered parton to a projectile nucleon following hard Pomeron (color singlet) exchange (see Sec.\ XIII-C,D of Ref.~\cite{fragevo}). That mechanism naturally produces jet elongation in {\em one direction} on $\eta$ which would be symmetrized in an ensemble average. It is also an ``early time'' phenomenon capable of producing structure over large rapidity intervals.


\subsection{$\bf p_t$ dependence of correlated pairs}

In Fig. 5 (lower panel) of Ref.~\cite{mg} correlation data are shown for $p_t$ cuts which correspond to a {\em running integral from below} of the SS 2D peak on $p_t$. The data as presented give the impression that the most probable particle momentum in the SS 2D peak is 0.2 GeV/c, apparently supporting the imposed ``soft ridge'' terminology. However, the underlying spectrum of correlated particles corresponds to the {\em negative derivative} of the plotted data. The true most-probable momentum for those data is 1 GeV/c, consistent with spectrum hard components for \pp\cite{ppprd} and more-peripheral \auau collisions~\cite{hardspec}. Spectrum hard components are in turn quantitatively described as jet fragment distributions by the folding of a pQCD dijet spectrum with measured fragmentation functions~\cite{fragevo}. The term ``soft'' is misleading.

The $p_t$ trend of the SS peak azimuth width is described in Ref.~\cite{mg} in terms of radial flow: larger pair $p_t$ corresponds to smaller opening angle because a fluid cell has larger radial boost. But the same trend is expected from pQCD jets, with ``fluid cell'' translated to location in the fragmentation cascade descending from a parent parton. That is one of several examples where pQCD jet phenomenology is recast within a hydro scenario. Another is interpretation of low-$p_t$ jet fragment distributions in terms of radial flow~\cite{hardspec,fragevo}. Distinctions made between a ``thermal bulk'' at smaller $p_t$ and jet fragments at larger $p_t$ are unjustified, since according to pQCD (MLLA~\cite{mlla}) most parton fragments appear below 2 GeV/c for any jet~\cite{eeprd,hardspec,fragevo,nohydro}, also implying that most jet-correlated hadron pairs appear at larger angular separations.

\subsection{Spectrum and correlation references}

For differential spectrum and correlation analysis of \aa collisions it is essential to define a Glauber linear superposition (GLS) reference based on differential measurements in elementary collisions. The GLS reference for two-component spectrum analysis is~\cite{hardspec,nohydro} 
\bea
\frac{2}{N_{part}}\frac{1}{2\pi y_t}\frac{d^2n_{ch}}{dy_td\eta} &=& S_{NN}(y_t) + \nu H_{NN}(y_t)
\eea
reflecting binary-collision scaling of fixed hard component $H_{NN}$. In Ref.~\cite{daugherity} the dashed correlation reference curve in Fig.\ 3 (left panel) represents  the GLS reference
\bea
 \frac{0.045 N_{bin}}{\frac{N_{part}}{2}[S_{NN} + \nu H_{AA}(b)] / \rho_{pp}} \approx \frac{0.045\, \nu}{1 + x(b) (\nu - 1)}
\eea
extrapolated from the SS 2D peak in 200 GeV \pp collisions, where $x(b)$ increases linearly from 0.02 in \pp collisions to $x \sim 0.1$ in central \auau collisions. $x = 0.02 \approx H_{NN} / (S_{NN} + H_{NN})$ in acceptance $\Delta \eta = 2$ is derived from a two-component analysis of \pp $p_t$ spectra~\cite{ppprd}. $x \sim 0.1$ matches spectrum yields in more-central \mbox{Au-Au}.

In Ref.~\cite{mg} similar dashed curves in Fig. 3 are described as ``wounded nucleons + flow.'' ``Flow'' should play no role in extrapolations from \pp collisions. The preferred (GLS) reference is based on known {\em binary-collision} scaling of measured jet correlations~\cite{porter1,porter2} and spectrum trends~\cite{ppprd} in \pp collisions.

 \section{Summary}

A saturation-scale argument has been proposed to explain the origin of the $\eta$-elongated same-side (SS) 2D peak (``soft ridge'') in $p_t$-integral (minimum-bias) angular correlations  in terms of Glasma flux tubes plus radial flow. The SS peak is consistent with expected number and $p_t$ angular correlations from jets in more-peripheral \aa and \pp collisions. The $\eta$-elongation phenomenon has been observed in more-central \auau collisions.

In the present analysis we have reviewed saturation-scale predictions for particle production and find that the prediction $\propto \log(8\nu)$ ($\nu = 2 N_{part} / N_{bin}$ measures the number of binary \nn collisions per participant pair) disagrees with measured hadron production, whereas a two-component pQCD analysis of jet angular correlations describes the hadron production data quantitatively.

We have examined a statistical argument relating the saturation scale to the amplitude of the SS 2D peak and find that two contributions to number fluctuations from angular correlation structure have been confused. The contribution which might correspond directly to the volume of the SS 2D peak (intra-source correlations) is assumed to be negligible, whereas a contribution which corresponds to a uniform correlation background is attributed to the SS peak amplitude.
%

In comparisons of predicted flux-tube correlation structure with measured 2D angular correlations we find that the predicted peak shape is excluded by the 2D correlation data for \pp and more-peripheral \aa collisions (small upper limit consistent with zero). Given the flux-tube centrality trend $\propto \log(8\nu)$ for the SS peak amplitude (the same trend describing particle production) the predicted SS correlation amplitude for more-central \aa collisions is much smaller than what is observed.  

Finally, we have compared features of Glasma flux tubes plus radial flow with pQCD jets as predictors of spectra and correlations. We find that whereas the pQCD jet description accommodates RHIC data quantitatively and comprehensively the Glasma flux-tube hypothesis disagrees with data in several aspects.

This work was supported in part by the Office of Science of the U.S. DOE under grant DE-FG03-97ER41020.


\end{document}